\documentclass[showpacs,preprintnumbers,amsmath,amssymb,eqsecnum]{revtex4}
\usepackage{graphicx}
\begin{document}
\preprint{IMAFF-RCA-07-08}
\title{A graceful multiversal link of particle physics to cosmology}

\author{Pedro F. Gonz\'{a}lez-D\'{\i}az$^{1}$ Prado Mart\'{\i}n Moruno$^{1}$ and Artyom V. Yurov$^{2}$}
\affiliation{$^{1}$Colina de los Chopos, Centro de F\'{\i}sica
``Miguel A.
Catal\'{a}n'', Instituto de Matem\'{a}ticas y F\'{\i}sica Fundamental,\\
Consejo Superior de Investigaciones Cient\'{\i}ficas, Serrano 121,
28006 Madrid (SPAIN).} \affiliation{$^{2}$I. Kant Russian State
University, Theoretical Physics Department 14, Al.Nevsky St.,
Kaliningrad 236041, Russia}
\date{\today}
\begin{abstract}
In this paper we work out a multiverse scenario whose physical
characteristics enable us to advance the following conjecture:
whereas the physics of particles and fields is confined to live in
the realm of the whole multiverse formed by finite-time single
universes, that for our observable universe must be confined just in
one of the infinite number of universes of the multiverse when such
a universe is consistently referred to an infinite cosmic time. If
this conjecture is adopted then some current fundamental problems
that appear when one tries to make compatible particle physics and
cosmology- such as that for the cosmological constant, the arrow of
time and the existence of a finite proper size of the event horizon-
can be solved.
\end{abstract}


\vspace{1.5cm}

\maketitle


\section{Introduction.}
The perception that the very big and the very small are both
governed by the same physical laws is an ancient conception that has
rendered extremely fruitful results such as the Galileo's mechanics
or the Newton's universal gravitation law. That conception actually
is a very popular one and is felt by many people at all cultural
levels. However, the current status of particle physics and
cosmology seems to inexorably avoid such a common treatment. In
fact, by the time being, in spite of the the success of the
inflationary paradigm, they look irreconcilable indeed [1]. There
are several main points for that discrepancy. In particular, the
known problem of the cosmological constant [2] by which the
predicted value of the quantum-field vacuum energy density and that
for cosmology are currently separated by many orders of magnitude,
the feature that whereas fundamental physical theories are
time-symmetric, cosmology contains an intrinsic arrow of time [3],
and the existence of a future event horizon with finite proper size
in accelerating cosmology which is mathematically and physically
incompatible with any fundamental theory, such as string theory or
quantum gravity, based on the introduction of an S-matrix requiring
the propagation between points infinitely space-separated [4].
Actually, that can be regarded to be one of the greatest problems of
all theoretical physics.

In this paper we take the above difficulties as being fundamental
and essentially inescapable, provided that one keep oneself and the
physics of particles and cosmology within the realm of a single
universe like the one which we live in. Really, what we are going to
argue is that all these difficulties simply vanish whenever we take
for cosmology a multiversal conception in such a way that whereas
fundamental physics lives in any of the finite-time universes of the
multiverse, our observed cosmology can only be described in one
single universe endowed with an infinite cosmic time, isolated out
from the multiverse and whose physical characteristics are precisely
and consistently relatable to those of the original whole
multiverse.

There already are a plethora of multiverse models, essentially
including those coming from quantum mechanics [5], those described
in inflationary theory [6], those which come about in string or M
theories [7], and those which are just based on classical general
relativity [8]. In order for trying to implement our main idea we
should choose a classical multiverse model that is able to account
for the current accelerating cosmic behaviour, leaving all quantum
considerations to be built up later on. One of such multiverse
models, which moreover, provides us with possibly the most general
framework, is the recently suggested dark energy multiverse [9].
Throughout this paper we shall make all our considerations in the
realm of such a model.

This idea of the multiverse and its relation with the physics in a
single universe reminds us of Plato's cave [10]. In this myth a
group of humans chained in a cave with their back to the entrance
with a bonfire between them and this entrance, thus they are
condemned to see only the shadows of the world outside. So Plato
explains the relation between the sensitive world, which we
perceive, and the world of ideas, the real world. Independently of
metaphysical considerations related to pure ideas and their
hierarchy, we could adapt this philosophical myth to a physical
myth, in such a way that the exterior of the cave is the true nature
which we can measure only through its shadows. Thus the real world
would play the role of the multiverse, where the particle physics
are well defined, but the world that we are able to experience, the
cosmological world, is only made of shadows projected from the real
world. So the apparent inconsistencies that could appear when
comparing the physics of particles and cosmology could just be seen
as artifacts arising from trying to identify shadows with real
world.

This paper can be outlined as follows. In Sec. II we briefly review
the model of dark energy multiverse recently suggested, emphasizing
the main points in that model which are going to be useful to
sustain the ideas and conjectures put forwards in the paper. The way
in which one can single out one of the infinite universes in the
multiverse and consistently convert it into our own universe is
described in some detail in Sec. III, where the vacuum structure and
the properties of the cosmic time for such an individualized
universe are also discussed. Sec. IV contains some calculations and
discussion that allow us to finally make the main conjecture of the
present work, that by choosing our own universe as just a single
part described in an infinite cosmic time of the whole multiverse
one may reconcile in a consistent way fundamental physics with
cosmology. We conclude and add some further comments in Sec. V. Two
appendices are added. In the first Appendix A we derive a
two-parameter generalization of the dark energy multiverse, and in
Appendix B the accretion of the cosmic fluid by wormholes is studied
in detail both in the original multiverse and in the resulting
observable universe. The results obtained in the latter appendix are
used to discuss the problem of time-asymmetry in our universe.

\section{Dark energy multiverse.}

In this section we shall briefly review the dark energy multiverse,
a scenario based on accelerating cosmology which has been recently
introduced in Re. [9]. If we consider a cosmological model which is
dominated by a quintessential fluid with constant equation of state
$p=w\rho=w\rho_0[a(t)/a_0]^{-3(1+w)}$ (where $a(t)$ is the scale
factor for the universe, and $p$ and $\rho$ are respectively the
pressure and energy density, with the subscript ``0'' denoting
current value) plus a negative cosmological constant, $\Lambda$,
then the Friedmann equation can be written as
\begin{equation}\label{uno}
H^2=-\lambda+Ca^{-3\beta},
\end{equation}
where $\lambda=|\Lambda|/3$, $C=8\pi\rho_0/(3a_0^{-3\beta})$ and
$\beta=1+w$. By integrating Eq.~(\ref{uno}), we can obtain the time
evolution of the cosmic scale factor, for $\beta<0$,
\begin{equation}\label{dos}
a(t)=a_0\left[\cos\left(\alpha(t-t_0)\right)-
b\sin\left(\alpha(t-t_0)\right)\right]^{-\frac{2}{3|\beta|}},
\end{equation}
with $\alpha=\frac{3|\beta|}{2}\lambda^{1/2}$ and
$b=\left(\frac{C}{\lambda}a_0^{-3\beta}-1\right)^{1/2}=
\left(\frac{8\pi}{3\lambda}\rho_0-1\right)^{1/2}$. Therefore, an
infinite number of big rip-like singularities will occur in this
model at times given by
\begin{equation}\label{tres}
t_{br_{m}}= t_0+\frac{2}{3|\beta|\lambda^{1/2}}{\rm
arctg}\left[\left(\frac{8\pi\rho_0}{3\lambda}-
1\right)^{-1/2}\right]+\frac{2m\pi}{3|\beta|\lambda^{1/2}},
\end{equation}
in which $m$ is any natural number. It is easy to check that, for
$m=0$, expanding the above expression for $\lambda<<1$, we recover
the occurrence time of the big rip for a quintessence model of
phantom energy without any cosmological constant, that is,
\begin{equation}\label{cuatro}
t_{br}=t_0+\frac{1}{|\beta|\left(6\pi\rho_0\right)^{1/2}}.
\end{equation}

Even though we have used a negative cosmological constant, the fact
that one can define an overall positive vacuum energy density has
led us to a true cosmic model. It is worth noticing that in order to
make all possible physical regions in the above solution physically
meaningful, the parameter of the equation of state should be
discretized, $|\beta|=\frac{1}{3n}$, with $n=1,2,3,...$, so
guaranteeing the scale factor to be always positive.

Due to the singular character of the big rips, in the absence of any
wormhole-type connection between single universes, the regions
between two such singularities are causally disconnected and each of
these regions can be interpreted as a different spacetime (see Ref
[9]), in fact a different universe within the whole infinite
multiverse. Classically and in the absence of observable matter, all
of these universes in the multiverse are physically
indistinguishable; all starting with an infinite size, which will
then steadily decrease until a minimum nonzero value,
\begin{equation}\label{cinco}
a_{{\rm min}}=a_0
\left(\frac{8\pi\rho_0}{3\lambda}\right)^{-1/3|\beta|}>0  .
\end{equation}
After that moment the universe acceleratingly expands again to
infinity.

We finally want to remark that even though all the single universes
in the multiverse are classically identical, it could well be that
one might envisage the contents of the set of universes like given
realizations of the quantum superposition in a quantum-mechanical
treatment, somehow parallelling the Everett's many-world
interpretation [11].

It is also worth noticing that a multiverse scenario with
essentially the same structure as the one discussed above can be
considered as well in the Randall-Sundrum brane with positive
tension, $\mu >0$, provided $2\mu >\lambda$. In fact, if the brane
is filled with phantom energy with equation of state $p=w\rho$
($w=-1-\alpha/3$, $\alpha>0$) and a negative cosmological constant,
$\Lambda<0$, such that $\lambda=-\Lambda/3 > 0$, then from the
Friedmann equations for the brane [12],
\begin{equation}
H^2 =\rho\left(1+\frac{\rho}{2\mu}\right)
\end{equation}
\begin{equation}
\dot{\rho}=-3H\left(\rho+p\right) ,
\end{equation}
we can obtain
\begin{equation}
\rho=-\lambda +Da^{\alpha},
\end{equation}
with $D$ a constant, and for $2\mu>\lambda$
\begin{equation}
a(t)^{\alpha}=\frac{\lambda(2\mu-
\lambda)}{D\left[2\mu\cos^2\left(\frac{\alpha\sqrt{\lambda(2\mu-
\lambda)}}{2\sqrt{2\mu}} t\right) -\lambda\right]} .
\end{equation}
It can be now immediately seen that there will be an infinite number
of big rip singularities at
\begin{equation}
t_{br}=
\frac{2\sqrt{2\mu}}{\alpha\sqrt{\lambda(2\lambda-\lambda)}}\left(\pm
{\rm arccos}\sqrt{\frac{\lambda}{2\mu}} +2\pi m\right) ,
\end{equation}
or
\begin{equation}
t_{br}=
\frac{2\sqrt{2\mu}}{\alpha\sqrt{\lambda(2\lambda-\lambda)}}\left(\mp
{\rm arccos}\sqrt{\frac{\lambda}{2\mu}} +2\pi(2m\pm 1)\right) .
\end{equation}
Thus, one can construct a multiversal scenario fully analogous to
the one discussed above for brane worlds with $\mu > \lambda/2$.
This will be no longer the case however if $\lambda\geq 2\mu$. In
fact, when $\lambda > 2\mu$, we get the solution
\begin{equation}
a(t)^{\alpha}=\frac{\lambda(2\mu-
\lambda)}{D\left[2\mu\sinh^2\left(\frac{\alpha\sqrt{\lambda(2\mu-
\lambda)}}{2\sqrt{2\mu}} t\right) +\lambda\right]} ,
\end{equation}
which does not show any big rip singularities and therefore cannot
be cut off in an infinite set of independent universes, and if
$\lambda=2\mu$ such a solution reduces to
\begin{equation}
a(t)^{\alpha} =\frac{4\lambda}{D\left(4-\lambda\alpha^2 t^2\right)}
\end{equation}
that describes a single universe which starts and dies at big rip
singularities taking place at
\begin{equation}
t_{br}=\pm\frac{2}{\alpha\sqrt{\lambda}} .
\end{equation}
Thus, the last two cases could not lead to any multiverse scenarios.

In the rest of the paper we shall restrict ourselves to the
braneless multiverse case (a generalized version of which is dealt
with in Appendix A), leaving the treatment of the brane multiverse
to be dealt with elsewhere. We only advance here that the brane
world topological defect splits in an infinite set of single
defects out from which just one may develop the physical
properties that we can observe in our universe.

\section{The observable single universe.}

Singling out a universe out from the whole multiverse should imply
the consideration of observers in that single universe that would in
principle interpret their spacetime as the unique, full spacetime
for which, in the absence of past or future singularities or
crunches, the time ought to be infinite. Since every single universe
in the multiverse is defined for a finite time interval, our first
task must be to re-scale the finite time interval of one such single
spacetimes so that it became infinite.

Thus, out from Eq.~(\ref{dos}) we first consider a single
finite-time universe whose scale factor is expressed as
\begin{equation}\label{seis}
a(\tau)=a_{{\rm min}}\cos^{-\frac{2}{3|\beta|}}(\tau),
\end{equation}
where $-\pi/2\leq\tau\leq\pi/2$ with
$\tau=\frac{3|\beta|}{2}\lambda^{1/2}(t-t_0)+{\rm
arctan}\left(\frac{H_0}{\lambda^{1/2}}\right)$, so that $a(\tau)$
reaches its minimum value at $\tau=0$. We can then refer the scale
factor to an infinite time interval by re-defining the time $\tau$
so that the new infinite time is given by $T=\tan(\tau)$, with
$-\infty\leq T\leq\infty$. As expressed in terms of time $T$ the
scale factor would become
\begin{equation}\label{siete}
a(T)=\frac{a_{{\rm min}}}{{\rm cos}^{2/(3|\beta|)}({\rm arctan} T)}=
a_{{\rm min}}(1+T^2)^{1/(3|\beta|)}.
\end{equation}

In order to obtain a more familiar expression for the scale factor,
it is convenient to re-define again $T$ in terms of another infinite
time given by
\begin{equation}\label{ocho}
\eta=\frac{1}{\lambda^{1/2}}{\rm arccosh}\left(1+T^2\right),
\end{equation}
so that
\begin{equation}\label{nueve}
a(\eta)=a_{{\rm min}}{\rm
cosh}^{1/3|\beta|}\left(\lambda^{1/2}\eta\right),
\end{equation}
with $-\infty\leq\eta\leq\infty$ again.

Thus, we have been able to derive an expression for the scale factor
which somehow resembles that for a de Sitter space. Even
Eq.~(\ref{nueve}) reduces to the scale factor for a de Sitter space
if we specialize to the case $n=1$, i. e. $|\beta|=1/3$. However, in
order to see how our model adjust to the available observational
data, one need to use the current value of the hyperbolic cosinus in
Eq.~(\ref{nueve}), whose expression in terms of $H_0$ and $\lambda$
can be derived by two distinct procedures: either by directly
specializing Eq.~(\ref{nueve}) to $\eta=\eta_0$, without any need of
differentiating $a(\eta)$ with respect to $\eta$, or by first using
the definition of $H$ in terms of the $\dot{a}(\eta)=da(\eta)/d\eta$
derivative and then specializing to $\eta=\eta_0$. Of course, for
the physical model to be consistent the above two procedures should
yield the same final expression. However, what we get instead is
${\rm cosh}\left(\lambda^{1/2}\eta_0\right)=(H_0^2+\lambda)/\lambda$
following the first procedure and ${\rm
cosh}\left(\lambda^{1/2}\eta_0\right)=\left(\frac{\lambda-
(3|\beta|)^2H_0^2}{\lambda}\right)^{-1/2}$, using the second one.
The ultimate reason for such a discrepancy should reside indeed in
the feature that the Friedmann equation~(\ref{uno}) describes the
whole multiverse, not every single isolated universe in it.
Actually, the most general Friedmann equation which is compatible
with a generic functional form for the scale factor like in
Eq.~(\ref{nueve}) can be checked to be
\begin{equation}\label{diez}
H^2=C_{{\rm n}}a^{-3\beta_{{\rm n}}}+\lambda_{{\rm n}},
\end{equation}
in which $C_{{\rm n}}=8\pi\rho_{{\rm n}0}a_0^{3\beta_{{\rm
n}}}/3<0$, $\beta_{{\rm n}}=1+w_{{\rm n}}>0$ and $\lambda_{{\rm
n}}=\Lambda_{{\rm n}}/3>0$, with $\rho_{{\rm n}0}$, $w_{{\rm n}}$
and $\Lambda_{{\rm n}}$ being general values to be specified later
on. Eq.~(\ref{diez}) admits the solution
\begin{equation}\label{once}
a(t_c)= a_{{\rm min}}{\rm cosh}^{\frac{2}{3\beta_{{\rm
n}}}}\left(\frac{3\beta_{{\rm n}}}{2}\lambda_{{\rm
n}}^{1/2}t_c\right),
\end{equation}
where $a_{{\rm min}}= a_0\left(\frac{8\pi|\rho_{{\rm
n}0}|}{3\lambda_{{\rm n}}}\right)^{\frac{1}{3\beta_{{\rm n}}}}$.

Equalizing finally Eqs.~(\ref{nueve}) and (\ref{once}) we get
\begin{equation}\label{doce}
\beta_{{\rm n}}=2|\beta|=\frac{2}{3n},
\end{equation}
\begin{equation}\label{trece}
\left(\frac{8\pi|\rho_{{\rm n}0}|}{3\lambda_{{\rm
n}}}\right)^{\frac{1}{3\beta_{{\rm n}}}}=
\left(\frac{8\pi\rho_0}{3\lambda}\right)^{-\frac{1}{3|\beta|}}
\end{equation}
and
\begin{equation}\label{catorce}
(\lambda)^{1/2}\eta=\frac{3\beta_{{\rm n}}}{2}(\lambda_{{\rm
n}})^{1/2}t_c.
\end{equation}
>From Eqs.~(\ref{doce}) and (\ref{trece}), using the definition of
$\lambda$ and $\lambda_{{\rm n}}$, we have
\begin{equation}\label{quince}
(8\pi)^3\frac{|\rho_{{\rm n}0}|}{\Lambda_{{\rm
n}}}=\frac{|\Lambda|^2}{\rho_0^2}.
\end{equation}

Now, the above two distinct procedures to check consistency of the
model produce the same expression for the hyperbolic cosinus as
referred to current time $t_{c0}$,
\begin{equation}\label{dieciseis}
{\rm cosh}^2\left(\frac{3\beta_{\rm
n}}{2}\lambda^{1/2}t_{c0}\right)=\frac{\lambda}{\lambda-H_0^2},
\end{equation}
so implying that the new time $t_c$ makes a well-defined choice for
the cosmic time.

Besides the above argument, full consistency of the model requires
that it produces a suitable acceleration; more precisely, we need
also that the expression for $\kappa=-q_0$ derived from the
Friedmann equation for a flat geometry is the same as that is
obtained by directly applying the definition of $q_0$ in the present
model, and that the predicted value of $\kappa$ be compatible with
the one which is expected for an accelerating universe, that is to
say [13], $\kappa$ should be slightly greater than unity
~\footnote{The WMAP 3-year data, in combination with large-scale
structure and SN data, allows for $w < -1$ in a general relativistic
model. For constant $w_0$ \cite{13} $w_0=-1.06^{+.13}_{-.8}$ and
$q_0=(1+3w_0)/2$. }. In fact, if we have $\Omega_T=\Omega_{\rm
n}+\Omega_{\Lambda_{\rm n}}=1$, with $\Omega=\rho/\rho_{{\rm
crit}}$, $\rho_{\Lambda_{\rm n}}=\Lambda_{\rm n}/\rho_{\rm n}$ and
$w_{\Lambda_{\rm n}}=-1$, and from the second Friedmann equation
\begin{equation}\label{diecisiete}
3\frac{\ddot{a}}{a}=-4\pi(3p_{T}+\rho_{T}),
\end{equation}
we can get
\begin{equation}\label{dieciocho}
\frac{\ddot{a_0}}{a_0}=H_0^2\left(\frac{3\beta}{2}|\Omega_{n}|+1\right),
\end{equation}
and whence
\begin{equation}\label{diecinueve}
\kappa=\frac{3\beta}{2}|\Omega_{n}|+1>1,
\end{equation}
which should in fact be just slightly greater than unity as $\beta$
and $|\Omega_{\rm n}|$ are both very small as we will see later on.

Now, directly differentiating the scale factor given by
Eq.~(\ref{once}) and using Eq.~(\ref{dieciseis}) we finally obtain
\begin{equation}
\kappa=\frac{3\beta}{2}\left(\frac{\lambda}{H_0^2}-1\right)+1,
\end{equation}
which can readily be seen to be the same as (\ref{diecinueve}).

Once we have checked the above consistency criteria, let us consider
the physics of the resulting cosmological model for one single
universe in the multiverse. Actually, after starting with a cosmic
model equipped with a negative cosmological constant plus a vacuum
phantom fluid characterized by a positive energy density and
$\beta<0$, so that the total vacuum energy density was positive, we
have finally singled out an observable universe which still has a
positive total vacuum energy density but now distributed as a sum of
a positive cosmological constant and a negative dynamical part. The
latter part corresponds to the so-called dual of dark energy, or in
short, dual dark energy [14]. It is generally defined as a fluid
having negative energy density and positive pressure, with
$\beta_{{\rm n}}$ taking on values from 0 to 2/3. Besides the
important property that the total energy density of the model is
definite positive, such a negative dynamic density-energy component
violates most energy conditions [15] and only may be allowed to
exist provided that it is very small, in a quantum-mechanical
context. This condition will be shown to be fulfilled in sec. IV
and, since the quantum inequality condition [16] that any existing
negative energy should be always accompanied by an overcompensating
amount of positive energy (here given by the cosmological constant
terms) is also satisfied, it appears that the resulting cosmic model
fulfills all observational requirements and can be taken to provide
us with a consistent and realistic scenario to deal with current
cosmology. In fact, such a scenario is again somehow similar to a de
Sitter framework and appears to also have a cosmological horizon.
Because $a(t_c)=a_{{\rm min}}{\rm cosh}^{n}\left(\frac{3\beta_{{\rm
n}}}{2}\lambda_{{\rm n}}^{1/2}t_c\right)$, where $n=1,2,3,...$ and
the case for $n=1$ corresponds to de Sitter universe, the
acceleration predicted by these models goes generally beyond that of
a de Sitter space, without giving rise to any future singularity of
the big rip type, a case which is certainly compatible with nowadays
observational data.

\section{Multiversal link between fundamental physics and cosmology}

In the Introduction it was pointed out that in spite of belonging to
a long and fruitful tradition the idea that the very large and the
very small are both governed by essentially the same laws has
reached a turning point during the last decades from which one only
finds failures in its application to current particle physics and
cosmology. Moreover, it is not just that such laws are not similar
or the same, but that these two branches of physics appear to be
actually incompatible. In this paper we distinguish three main
situations where the discrepancies are most apparent: the so-called
problem of the cosmological constant, the existence of an arrow of
time in cosmology and the current cosmological prediction that there
exists a future event horizon whose proper size is finite. In what
follows we shall argue that the headaches produced by these three
apparently basic difficulties all vanish in the multiverse framework
considered in this paper. We actually content that the above three
shortcomings are nothing but artifacts coming from considering as
the universe what really is nothing but a part of the whole physical
reality. That is to say, whereas the fundamental physics resides and
is well-defined in the whole realm of the multiverse (or just in one
of its finite-time universes if all the multiversal components are
identical), what we usually take as cosmology is defined just for
one of the infinite independent spacetimes which the multiverse is
made of when it is referred to a suitable infinite cosmic time.

In order to satisfy the observational data it is necessary that
$H_0^2\sim \lambda_n\sim 10^{-52} {\rm m}^{-2}$ [17]. Furthermore,
the Friedmann equation for the single infinite-time universe,
Eq.~(\ref{diez}), imposes that $\lambda_n > H_0^2$ as $C_n$ is
definite negative. It follows that the absolute value of $\rho_n$
must be very small and therefore negative energies could only appear
in our observable universe when they are small enough. On the other
hand, if we consider that the absolute value of the constant vacuum
energy density in the multiverse corresponds to the Planck value,
i.e. $\rho_{\lambda}^{Pl}\sim 10^{110} {\rm erg}/{\rm cm}^3$, then
$\lambda\sim 10^{62} {\rm m}^{-2}$. Taking into account relation
(\ref{trece}) we then have
\begin{equation}
\rho_0\sim \frac{10^{35} {\rm m}^{-3}}{|\rho_{n0}|^{1/2}} .
\end{equation}
In addition, in order to have a positive definite Hubble parameter
also in the multiverse it is required that $\rho_0 >3\lambda/(8\pi)
\sim 10^{61} {\rm m}^{-2}$. From these considerations, it follows
that
\begin{equation}
|\rho_{n0}|\sim\frac{10^{70}}{\rho_0^2} < 10^{-52} {\rm m}^{-2} .
\end{equation}
Because this should be always satisfied, as we have pointed out
above, we can finally conclude that in the present scenario it is
natural to have a cosmological constant in the multiverse with a
value compatible with high energy physics and simultaneously a much
smaller value for that constant of the order of those predicted in
current cosmology in our observable universe.

Any cosmological models, including the one which gives rise to the
multiverse, possesses an intrinsic arrow of time, that is a
privileged direction along which the time only flows towards the
future. If the physics of particles and fields is time symmetric and
should be described in the realm of the whole multiverse (or just in
one of its finite-time universes if all the multiversal components
are identical), then one must have a deep physical reason that makes
the microscopic behaviour to appear as time symmetric. Such a
physical reason can be found if we consider the existence of
wormholes in the realm of the multiverse. Since these wormholes
accrete dark energy [18] they can actually grow so big that the
whole spacetime of any of the universes making the multiverse is
engulfed by the wormhole immediately before (after) the universe
reaches (leaves) the future (past) big rip singularity. Such a
gigantic process has been denoted as big trip and has hitherto been
considered to just predict unwanted catastrophic cataclysms in the
future of a hypothetical universe filled with phantom energy [18].
Nevertheless, when considered in the context of our multiversal
model, the phenomenon of the big trip may provide unexpected
benefits. If fact, it can be shown (see Appendix B) that an observer
in one of the finite-time universes that make the infinite
multiverse will see big trips to crop up in his (her) future and his
(her) past. Thus, such as it was pointed out in Ref.~[14], the
mouths of the wormholes in the past and future may be moving, and so
travelling in time, in such a way that they can be inserted into
each other during each big trip time so that the given universe can
freely journey from future to past and vice verse, so destroying any
arrow of time of that universe and rendering the time fully time
symmetric in the whole multiverse. In Appendix B it is also shown
that any big trip phenomena are prevented to take place in a single
universe with infinite time, and therefore a reason is found why
current cosmology keeps an arrow of time.

The cropping up of wormholes with distinct sizes in the neighborhood
of the big rips [19] of the multiverse helps us to furthermore
consider a future event horizon for any observer in the multiverse,
defined to have a proper size given by
\begin{equation}
R_h = a(t)\int_t^{\infty}\frac{dt'}{a(t')} .
\end{equation}
Inserting then the expression for the scale factor given by
Eq.~(\ref{dos}) one can easily check [20] that $R_h$ becomes
infinite for any observer in the multiverse, so making
mathematically and physically fully consistent the consideration of
any fundamental theory based on the definition of an S-matrix in the
context of the multiverse. Since a single universe with infinite
time resembles a de Sitter space and hence contains a future event
horizon with finite proper size, it is not possible to define such
fundamental theories in what we now consider as cosmology.

Based on the above considerations we can conjecture that the whole
physical reality consists of a multiverse whose structure can by
instance be described by the model reviewed in Sec. II, which is a
natural framework to consistently describe all time-symmetric
particle physics and fields, being what we call our own universe
just one among the infinite number of independent, identical
universes that form up the multiverse, whenever it is referred to a
consistently extended to infinity cosmic time. Such a singled
component is perceived by observers in it as a space which currently
expands in an super-accelerated fashion along an infinite cosmic
irreversible time in a similar manner to as de Sitter space does. If
that conjectured description is adopted, then the known
incompatibilities between particle physics and cosmology fade out.
The price to be paid is having a cosmological model for a fluid
characterized by a dynamics negative energy density (which is
nevertheless over-compensated by a positive cosmological constant in
accordance with the requirements of quantum theory) which is defined
on a physical domain where the quantization rules should be expected
to be not the same as those used in the whole multiverse (or just in
one of its finite-time universes if all the multiversal components
are identical) for particles and fields.

\section{Conclusions and further comments.}

We have tried to solve the incompatibilities between particle
physics and cosmology by resorting to the Copernican principle that
every founded scientific advance must necessarily be accompanied by
lowering of the human role in nature in the following sense. The
universe which we live in is nothing but one among the infinite
number of universes in a whole multiversal scenario. If we adscribe
to such a spacetime the observable properties of our own universe as
referred to an infinite cosmic time, then we derived in this paper a
cosmic model in which we can provide with some tentative solutions
to several key problems that arise when one tries to make compatible
particle physics with current cosmology.

Starting with a multiverse model recently suggested, we have singled
out a universe whose scale factor has been derived in terms of an
infinite cosmic time. That universe is interpreted as being ours own
and is characterized by a total positive vacuum energy density which
is made of two parts, a dynamical one which is small and negative
and a positive cosmological constant. We have then been able to
establish precise mathematical relations between the cosmological
parameters of the multiverse and those of the observable universe
which actually restore compatibility between cosmology and particle
physics if the latter is taken to be defined in the original
multiverse (or just in one of its finite-time universes if all the
multiversal components are identical). Thus, we have shown that: (i)
the ratio of values of the cosmological constant for particle
physics and cosmology derived in this way is precisely what has been
considered as the basis to formulate the so-called problem of the
cosmological constant, (ii) because for an observer in any of the
finite-time universes of the multiverse there are two big trip
phenomena, one in the past and other in the future, all the physics
in the multiverse will be time symmetry, but as no such phenomena
may take place in the single infinite-time universe, there will be
an arrow of time in our observable universe, and (iii) the proper
size of the event horizon of the multiverse is seen to be infinity,
i. e., there is no future event horizon in the multiverse, and
therefore any fundamental physics description can be consistently
carried out. All the above results amount to our main conjecture: If
we assume that particle physics lives in the multiverse (or just in
one of its finite-time universes if all the multiversal components
are identical) where it is well defined, then our cosmology results
from the observation referred to an infinite cosmic time of just one
of the infinite number of universes that form up the multiverse,
with the known incompatibilities between particle physics and
cosmology turning out to be nothing but artifacts arising from our
attempt to interpret our own universe as containing everything.

Obviously a more realistic model should require the introduction of
some matter. It is easy to see that if one add a matter term to the
Friedmann equation (\ref{uno}), this term would dominate only at
small values of the scale factor, becoming negligible on large
values of the scale factor. Thus, our multiverse scenario arising
from the occurrence of an infinite number of big rip singularities
at which the size of the universes blows up would still be valid.

\acknowledgements

\noindent This research was supported by MEC under Research Project
No. FIS2005-01181. The authors thanks C.L. Sig\"{u}enza for useful
discussions. PMM gratefully acknowledges the financial support
provided by the I3P framework of CSIC and the European Social Fund.

\appendix

\section{}

We derive here a two-parameter cosmic solution which shows the
properties of a multiverse that is the generalized form of the case
reviewed in Sec. II. Let us consider a universe filled with a
negative cosmological constant which, for the sake of convenience,
we denote now as $-\lambda^2$, and phantom energy with density
$\rho_{ph}=E^2 a^{\epsilon}$, with $E^2$ a constant and $\epsilon
>0$. Integrating then the Friedmann equation for a state equation
parameter $w=-1-\epsilon/3$ we have
\begin{equation}
a(t)=\left(\frac{\lambda}{E\cos\omega}\right)^{2/\epsilon},\;\;
w=\frac{\lambda\epsilon}{2}\left(t-W\right) ,
\end{equation}
in which $W$ is an arbitrary constant. A generalized dark energy
multiverse can then be derived by resorting to the following theorem
[21]. Let $a=a(t)$ be a spatially flat solution of the Friedmann
equations for energy density and pressure as given by
\begin{equation}
\rho=\frac{\dot{a}^{2}}{a^{2}} ,\;\;\;
p=-\frac{2a\ddot{a}+\dot{a}^{2}}{3a^{2}} .
\end{equation}
It follows then that the two-parameter function $a_k=a_k\left(t;c_1
, c_2\right)$, that is
\begin{equation}
a_k=a\left(c_1 +c_2\int\frac{dt}{a^{2k}}\right)^{1/k}
\end{equation}
will also be a solution of the Friedmann equations for new
expressions of the energy density, $\rho_k$, and pressure, $p_k$,
satisfying
\begin{equation}
k^2\rho_k -\frac{3k}{2}\left(\rho_k +p_k\right) = k^2\rho -
\frac{3k}{2}\left(\rho +p\right) ,
\end{equation}
This theorem implies that the quantity $k\left[(k-1)\dot{a}^2
+a\ddot{a}\right]/(a^2)$ is invariant under the transformation
$a\rightarrow a_k$, with $a_k$ as defined by Eq. (A3). Taking then
$k=\left(2m+1\right)\epsilon/4$, $m=0,1,2,...$, and using Eq. (A3),
one can get
\begin{equation}
a^{(m)}=
\left(\frac{\lambda}{E\cos\omega}\right)^{\frac{2}{\epsilon}}
\times\left[c_1 + c_2\sum_{s=0}^{m}\frac{(-1)^s}{2s+
1}C^s_m(\sin\omega)^{2s+1}\right]^{\frac{4}{2m+1}\epsilon} ,
\end{equation}
which describes a dark energy multiverse with a two-parameter
freedom. This is the wanted generalization of the dark energy
multiverse described in Sec. II to which it reduces when we take
$c_1=1$ and $c_2=0$.

\section{}

We consider in this appendix the accretion of the cosmic fluid onto
wormholes in both the multiverse and the single universe with an
infinite time [18] and [22]. It is known that the wormhole mass rate
for an asymptotic observer is expressed through
\begin{equation}\label{seisa}
\dot{m}=4\pi m^2Q|\beta|\rho.
\end{equation}
In first place, we study this process in the multiverse scenario
where the above expression must be integrated taking into account
Eq.~(\ref{dos}) and leads to
\begin{equation}\label{sietea}
m(t)= m_0\left[1 -\frac{8\pi
Q\rho_0m_0}{3\lambda^{1/2}}\frac{\sin(\alpha(t-t_0))}{\cos(\alpha(t-t_0))-
b\sin(\alpha(t-t_0))}\right]^{-1}.
\end{equation}
One can easily see that this expression vanish at the big rip times
and that it diverges an infinite number of times at
\begin{equation}\label{ochoa}
t_{*_{m}}=t_0+\frac{2}{3|\beta|\lambda^{1/2}}{\rm
arctan}\left[\frac{1}{b+
\xi}\right]+\frac{2m\pi}{3|\beta|\lambda^{1/2}},
\end{equation}
where $\xi=\frac{8\pi Q\rho_0m_0}{3\lambda^{1/2}}$ and $m$ is an
integer number. As in the big rip case, for $m=0$ and expanding
Eq.~(\ref{ochoa}) for $\lambda<<1$, one can recover the big trip
time expression of a simple phantom quintessential model, that is
\begin{equation}\label{nuevea}
t_{bt}=t_0+\frac{t_{br}-t_0}{1+\left(\frac{8\pi\rho_0}{3}\right)^{1/2}Qm_0}.
\end{equation}

In order to find the divergences of the wormhole mass, as expressed
through Eq.~(\ref{ochoa}), we must consider two cases,
$t_{*_{m}}-t_0<0$ and $t_{*_{m}}-t_0<0$. In the first case,
$t_{*_{m}}-t_0<0$, if we take into account Eqs.~(\ref{tres}) and
(\ref{ochoa}), then one obtains
\begin{equation}\label{dieza}
t_{br_{m}}-t_{*_{m}}= \frac{2}{3|\beta|\lambda^{1/2}}\left[{\rm
arctg}\left(\frac{1}{b}\right)-{\rm
arctg}\left(\frac{1}{b+\xi}\right)\right]>0.
\end{equation}
Since the divergences take place before the consecutive big rip,
this divergence will be a big trip (it can be seen that that is
actually the case if $t_{*_{m}},t_0>0$ and $t_{*_{m}}>t_0$, if
$t_{*_{m}},t_0<0$ and $|t_0|>|t_{*_{m}}|$ and if $t_{*_{m}}>0$ and
$t_0<0$). It can also be seen that the time interval between two
consecutive big rips is the same as between two big trips.

In the second case, $t_{*m}-t_0<0$, Eq.~(\ref{dieza}) is converted
into
\begin{equation}
t_{br_{m}}-t_{*_{m}}= \frac{2}{3|\beta|\lambda^{1/2}}\left[{\rm
arctg}\left(\frac{1}{b}\right)-{\rm
arctg}\left(\frac{1}{b+\xi}\right)-\pi\right]<0,
\end{equation}
where, because $t_*<t_0$, we have considered the previous big rip.
Then, an observer will see a big trip in his (her) past taking place
after the big rip singularity which is the origin of his (her)
universe (that is the case if $t_{*_{m}},t_0>0$ and $t_0>t_{*_{m}}$,
if $t_{*_{m}},t_0<0$ and $|t_0|<|t_{*_{m}}|$ and if $t_{*_{m}}>0$
and $t_0>0$). In short, for every observer in every finite-time
universe, there will be one big trip in the past and one big trip in
the future.

On the other hand, if we consider an observer which lives in a
single universe described in terms of a suitable infinite cosmic
time, as we should in fact be, we must take into account
Eq.~(\ref{once}) in order to integrate Eq.~(\ref{seisa}). So we get
\begin{equation}\label{treintauno}
m(t_c)=m_0\left[1+Am_0H_0-Am_0\lambda_{{\rm n}}^{1/2}{\rm
tanh}\left(\frac{3\beta_{{\rm n}}}{2}\lambda_{{\rm
n}}^{1/2}t_c\right)\right]^{-1}.
\end{equation}
We can define now the function $F(t_c)= Am_0\lambda_{{\rm
n}}^{1/2}{\rm tanh}\left(\frac{3\beta_{{\rm n}}}{2}\lambda_{{\rm
n}}^{1/2}t_c\right)$. Because of the properties of the tanh, we can
see that $F(t_c)$ is a monotonous increasing function, $F(t_c)>0$ if
$t_c>0$ and $F(t_c)<0$ if $t_c<0$, and $-Am_0\lambda_{{\rm
n}}^{1/2}<F(t_c)<Am_0\lambda_{{\rm n}}^{1/2}$. So, one has a zero in
the denominator of Eq.~(\ref{treintauno}) where the wormhole mass
would diverge if and only if $Am_0\lambda_{{\rm
n}}^{1/2}>1+Am_0H_0$, i. e., $m_0>1/[A(\lambda_{{\rm
n}}^{1/2}-H_0)]$. By mere inspection of the data in the last
section, it can be readily seen that this value of $m_0$ corresponds
to a minimum throat of order $b_{{\rm min}}\sim10^{26}({\rm
meters})$. However, if we assume that the Universe is expanding in
size at the speed of light, then its radius would be 13.7 billion
light years, and its diameter would be 27.4 billion light years,
which in meters is of the order $10^{26}$. Then, to have a wormhole
within the universe which could produce a big trip phenomenon, the
throat of this wormhole must be at less so big as the observable
universe, i. e., the universe would be already contained within a
wormhole. Since that situation is not possible, we can conclude that
a big trip phenomenon cannot take place in our single observable
universe.

\end{document}